\documentclass[prl,preprint,showpacs,preprintnumbers,floatfix]{revtex4}

\usepackage{epsfig}
\usepackage{graphicx}
\usepackage{dcolumn}
\usepackage{bm}

\newcommand{\beq}{\begin{eqnarray}}
\newcommand{\eeq}{\end{eqnarray}}
\newcommand{\bfig}{\begin{figure}}
\newcommand{\efig}{\end{figure}}
\newcommand{\btab}{\begin{table}}
\newcommand{\etab}{\end{table}}
\begin{document}
\title{Quantum Dynamics of Polaron Formation}
\author{Li-Chung\ Ku$^{1,2}$ and S.\ A.\ Trugman$^1$}
\address{
$^1$Theoretical Division, Los Alamos National Laboratory,
    Los Alamos, New Mexico 87545, U.S.A.\\
$^2$Department of Physics, University of California,
    Los Angeles, California 90024, U.S.A.}
\date{\today}

\begin{abstract}
The formation of a polaron quasiparticle from a bare electron
is studied in the framework of the Holstein
model of electron-phonon coupling.
Using Schr\"{o}dinger's formalism, we calculate the time evolution of
the distribution of the electron and phonon density, lattice deformation, and the
electron-phonon (el-ph) correlation functions in real space. The quantum
dynamical nature of the phonons is preserved. The polaron formation time is
related to the dephasing time of the continuum of unbound phonon excited
states in the spectral function, which depends on the phonon frequency and the
el-ph coupling strength. As the el-ph coupling increases, qualitative changes in
polaron formation occur when the one-phonon polaron bound state forms. In
the adiabatic regime, we find that a potential barrier between the quasi-free
and heavy polaron states exists in both 2D and 3D, which is crucial for polaron
formation dynamics.
 We compare to recent experiments.
\end{abstract}
 \maketitle


The polaron formation problem not only bears fundamental interest in
understanding polarization in condensed matter, but also determines the
electronic and optical properties in some applied materials, such as manganites
\cite{Salamon2001} and conductive organic oligomers \cite{campbell}. Recent
advances in ultra-fast time-resolved spectroscopy have made it possible to
investigate physical phenomena on the time scale of a molecular vibration or
optical phonon in crystals. Important aspects of chemical reaction
dynamics have been revealed by observation of the motion of the oscillating
atoms \cite{fs_book1}. Observations of the femtosecond dynamics of polaron
formation have recently been reported \cite{Ge98,Sugita,MX00,STE,averitt02}. In
contrast, theoretical development in this subject is somewhat behind. Although
the theoretical research on polaron physics
spans six decades \cite{alexandrov,devreese}, the dynamics of
lattice relaxation leading to a heavy polaron state remains far from being
understood. An adiabatic polaron theory \cite{adiab1} suggested that there is a
potential barrier between the delocalized and the self-trapped polaron states
in 3D but not in 1D or 2D. It is generally believed that the potential
barrier tends to impede the polaron formation. However, more careful thought is
needed to go beyond adiabatic limit (i.e.~$\omega_0 > 0$) where all
eigenstates are delocalized. It is not clear whether the barrier slows down
or forbids the formation of a polaron.

Measuring polaron formation time has been the subject of the
above recent experiments
\cite{Ge98,Sugita,MX00,STE,averitt02}. The time to form a polaron is found to
be less than a picosecond in strongly coupled systems, on the order of a phonon
period. Our goal is to provide a theory for the following fundamental
questions: (1) How are phonon excitations triggered and how do they evolve
into the correlated phonon cloud of the polaron quasiparticle? (2) How much
time does it take to form a polaron? (3) What is the effect of dimension on the
dynamics of polaron formation?


To understand the formation process of a polaron, we examine how the
bare particle wave function time evolves into a polaron quasiparticle. One
approach is to construct a variational many-body Hilbert space including
multiple phonon excitations, and to numerically integrate the many-body
Schr\"odinger Equation, $ i { {d \psi} \over {d t} } = H \psi $ in this space
\cite{Kenrow96}. Thus the full many-body wave function can be obtained
at early times. The main approximation is the size of the
variational space, which can be increased systematically until convergence is
achieved. This method includes the full quantum dynamics. Alternative
treatments, such as the semiclassical approximation \cite{semi_cl}, can be
inaccurate when applied to the present problem.
We solve for the dynamics of the Holstein Hamiltonian
\begin{eqnarray}
H &=& H_{el} + H_{\mbox{\scriptsize {\it el-ph}}} + H_{ph} \nonumber \\
  &=& - t\sum\limits_{<i,j>}  (c_i^\dag c_{j}  + h.c.) -
\lambda \sum\limits_j {c_j^\dag c_j (a_j  + a_j^\dag )} \nonumber \\ &~~& +~
\omega_0 \sum\limits_j {a_j^\dag a_j } ~,  \label{eq:ham}
\end{eqnarray}
where $c_j^\dag $ creates an electron and $a_j^\dag $ creates a phonon on site
$j$. The parameters are the nearest-neighbor hopping integral $t$, the el-ph
coupling strength $\lambda$, and the optical phonon frequency $\omega_0$.

Figure \ref{fig:snapshot} shows snapshots of polaron formation at weak
coupling. An initial bare electron wave packet is launched to the right as
shown in panel (a). (This initial condition is relevant to the recent
experiments \cite{Ge98,Sugita,MX00,STE,averitt02},
and to electron injection from a time-resolved STM
tip \cite{Toni2000}).
In panel (b) the electron is not yet dressed and thus is moving roughly as fast
as the free electron (dashed line). In addition, there exists a
back-scattering current (which evolves into a left-moving polaron later) on the
left side of the wave packet (green and black curves).
In panel (c) after an elapsed time of one phonon
period, the electron density consists of two peaks. The peak on the right
(black arrow) is essentially a bare electron. The peak on the left is a polaron
wave packet moving more slowly. The velocity operator is defined as $ V_j
\equiv {2J_{j,j+1} \over e(c_j^\dag c_j + c_{j+1}^\dag c_{j+1} )}$, where $j$ is
the site index and $J$ is the current operator $J_{j,j+1} = -i~et~(c_j^\dag
c_{j+1} - H.c.)~.$ $\langle V_j \rangle$ is shown in green.  As
time goes on, the bare electron peak decays and the polaron peak grows. Some
phonons are left behind (blue line), mainly near the injection point. These
phonons are of known phase with displacement
shown in red. Some phonon excitations travel with the
polaron (magenta). Finally, a coherent polaron wave packet is observed when
the polaron separates from the uncorrelated phonon excitations.


There are regimes where the polaron formation time is a calculable constant of
order unity times a phonon period $T_0$, as
seen in some experiments and shown in Fig.~\ref{fig:snapshot},
but there are other regimes where the phonon period is not the relevant
timescale. The limit hopping $t \rightarrow 0$ is instructive. After a time
$T_0 / 4$, the expectation of the phonon coordinate $\langle X_j \rangle$ on the
electron site has the same value as a static polaron.  It is tempting (but we
would argue incorrect) to identify this as the polaron formation time.  At
later times, $\langle X_j \rangle$ overshoots by a factor of two, and after
time $T_0$,  $\langle X_j \rangle$ and all other correlations are what they
were at time zero when the bare electron was injected. The system oscillates
forever.  In general an electron emits phonons enroute to becoming a polaron,
and we propose that the polaron formation time be defined as the time required
for the polaron to physically separate from the radiated phonons. The
polaron formation time for hopping $t \rightarrow 0$ is thus infinite, because
the electron is forever stuck on the same site as the radiated phonons.

An electron injected at several times the phonon energy $\omega _0$
above the bottom of the band is another instructive example. The electron
radiates successive phonons to reduce its kinetic energy to near the bottom of
the band, and then forms a polaron. For weak el-ph coupling $\lambda$, the rate
for radiating the first phonon can be computed by Fermi's golden rule, \beq
 \tau_{FGR}^{-1} = \frac{\lambda^2}{ \hbar~ t \sin(k_f)}, \label{eq:FGR}  \eeq
where $k_f$ is the electron momentum after emitting a phonon.  The
phonon emission time can be arbitrarily longer than the phonon period $T_0$ for
small $\lambda$. For strong coupling, the rate approaches $ \tau_{SC}^{-1}
= \lambda/\hbar $ because the spectral function smoothly spans numerous
narrow bands and its standard deviation is equal to $\lambda$. Our
numerical results agree with the perturbation theory (not shown).


We now consider polaron formation in more detail. After injecting a bare
electron at time zero, the wavefunction at later times $\tau$ is
 \beq
|\psi(\tau)\rangle = \sum_{j=1}^\infty e^{~-i E_j \tau}|\Psi_j\rangle \langle
\Psi_j | c^\dagger _ k |0\rangle , \label{eq:time_evolve} \eeq
where $|\Psi_j\rangle$ are a complete set of total momentum $k$ eigenstates of
the system of one electron coupled to phonons. There are several distinct types
of states contributing to the infinite sum: (A.1) The state $|k \rangle$ of a
momentum $k$ polaron, corresponding to the quasiparticle pole. (A.2) The states
$|k - q ; ~ q \rangle$ corresponding to a polaron of momentum $k - q$ and
an unbound phonon of momentum $q$, for any $q$. Similarly for two unbound
phonons, etc. If the electron-phonon coupling is sufficiently strong, there are
additional states in the sum: (B.1) A polaron excited state consisting of a
polaron and an additional bound phonon of total momentum $k$ \cite{bound_st},
designated $|k^{(I)} \rangle$.  This is a second type of (excited)
quasiparticle pole that is also split off from the continuum. (B.2) The states
$|(k - q)^{(I)} ; ~ q \rangle$ corresponding to an excited state polaron of
momentum $k - q$ and an unbound phonon of momentum $q$ for any $q$.
Similarly for two unbound phonons, etc.
For stronger el-ph coupling, more highly excited polaron states
corresponding to bound states of a polaron and two or more
additional phonons, $|k^{(II)} \rangle, \dots$
enter the sum \cite{otherstates}.
The branching ratios into the
various channels are calculated in Ref.~\cite{ku_thesis}.

From Eq.~\ref{eq:time_evolve}, the amplitude to remain in the initial state
after time $\tau$, $\langle \psi(\tau) | c^\dagger _ k |0\rangle$, is given by
the Fourier transform of the spectral function
\begin{eqnarray}
A(k, \omega) = \sum_{j=1}^\infty | \langle \Psi_j | c^\dagger _ k |0\rangle |^2
\delta(\omega - \omega _j ). \label{eq:spectral}
\end{eqnarray}
The numerically determined spectral function at fairly weak coupling is shown
in Fig.~\ref{fig:spc1}. We compute the spectral function by a Lanczos
algorithm. The sum rule $N( \omega )$ shows that about 10\% of the total
spectral weight is at energies beyond the range plotted.  There is a
quasiparticle pole corresponding to (A.1) above, and a group of states that is
approaching an approximately Lorenzian continuum as the number of sites
increases, corresponding to (A.2) above. The coupling $\lambda / \omega _0 =
0.8$ is too weak to form bound quasiparticle excited states $|k ^{(I)}
\rangle$. If the spectral function were a pure Lorenzian, a measurement would
yield a exponential decay of the initial state, with the polaron formation time
$\tau_p$ the inverse width of the Lorenzian.  Since the spectral function has a
quasiparticle pole as well, an experiment would measure the probability to
remain in the initial state
\begin{eqnarray}
P(\tau) = a_1^2 + a_2^2 e^{-2b\tau} + 2 a_1 a_2 e^{-b\tau} \cos [ ( \omega _1 -
\omega _2 ) \tau ] . \label{eq:prob}
\end{eqnarray}
This form already shows some complications, with an additive constant, a pure
exponential decay, and an exponential decay half as fast multiplied by a cosine
oscillating at the energy difference between the quasiparticle pole and the
center of the Lorenzian. Decaying oscillations in polaron formation (actually
the formally equivalent problem of an exciton coupled to phonons \cite{rashba})
have been observed in a pump-probe experiment \cite{Sugita} which measures
reflectivity after a bare exciton is created. The observed oscillatory
reflectivity was interpreted as the lattice motion in the phonon-dressed (or
self-trapped) exciton level. Assuming the modulation in the exciton level goes as
$\Delta E = -\lambda c_j^\dag c_j X_j$ where $X_j$ is lattice displacement, the
model Hamiltonian applies directly to the experiment. We calculate the
corresponding el-ph correlation function
$\chi \equiv \langle c_j^\dag c_j (a_j + a^\dag_j) \rangle$,
which is shown in Fig.~\ref{fig:Sugita}.
The result is in satisfactory agreement with experiment \cite{Sugita}.
(Numerical calculations in Fig.~\ref{fig:Sugita}
through Fig.~\ref{fig:levels_3D} are performed on an extended system, not a
finite cluster.)

Figure \ref{fig:spc2} shows the spectral function at stronger coupling than
Fig.~\ref{fig:spc1}. A polaron and two excited state polaron poles, along with
three continua are shown.  There is additional structure at higher energy (not
shown). The probability decay $P(\tau)$
for this spectrum is considerably more complicated, and
includes oscillating terms that do not decay to zero (at zero temperature) from
the ground and excited polaron poles beating against each other.


Next we discuss the role of dimensionality. 
The effect of dimensionality on static properties has been
studied previously \cite{pol_3d}.
The eigenvalues of the low-lying states (with
zero total momentum) are shown as functions of $\lambda$ in
Fig.~\ref{fig:levels_3D}. It is found that the energy spectra in $D>1$ are
qualitatively different than in $1D$. The 1D polaron ground state becomes
heavy gradually as $\lambda$ increases. However, in 2D or 3D, the ground state
crosses over to a heavy polaron state by a {\em narrow avoided level crossing},
which is consistent with the existence of a potential barrier \cite{adiab1}.
In the inset of
Fig.~\ref{fig:levels_3D}, $\psi_1$ and $\psi_4$ are
nearly free electron states;
$\psi_2$ and $\psi_3$ are heavy polaron states. The inner
product $|\langle \psi_1 | \psi_4 \rangle|$ is equal to 0.99. (A similar
avoided level crossing also occurs in 2D.)
Just right of the crossing region the effective
mass (approximately equal to the inverse of spectral weight) of the first
excited state can be smaller than the ground state by 2 or 3 orders of magnitude,
while their energies can differ by much less than $\omega_0$.
As a result, there is no optical phonon of the correct energy
for the light electron to emit and become a heavy polaron,
and the polaron formation time at $T=0$ would be infinite unless
low energy acoustic modes are included in the model.

In summary, we have calculated the time evolution of the many-body wave
function and find that the bare electron evolves into a
polaron quasiparticle by emitting phonons.
The excess kinetic energy is used to excite
uncorrelated phonons. The question ``how long
does it take a polaron to form?'' may not have a simple answer, given the
potentially complicated form of $P(\tau)$.  This function has been calculated
numerically, and at zero temperature depends on the parameters of the
Hamiltonian, the spatial dimension, the initial bare electron momentum $k$, the
final polaron momentum, and the possible existence of bound polaron
excited states. A further complication is that decay out of the initial state
need not be synonymous with decay into a polaron final state, as seen from the
weak coupling $\lambda$ regime, Eq.~\ref{eq:FGR}. In
addition, we show that a tunneling barrier between
the quasi-free and heavy polaron
state exists in both 2D and 3D (when $0< \omega_0 \ll 2Dt$). As a consequence,
the tunneling barrier inhibits the formation of a polaron in the crossover
regime. Our approach is intuitive and can be extended to other types of
quasiparticle formation, such as the vibrational exciton \cite{hamm} and spin
polaron.

This work was supported by the US DOE.
The authors thank A.~Alexandrov, I.~Bezel, A.~Bishop,
G.~Kalosakas, K.-K.~Loh, and A. Taylor for valuable discussions.


\newpage

\bfig  \centerline{ \psfig{figure=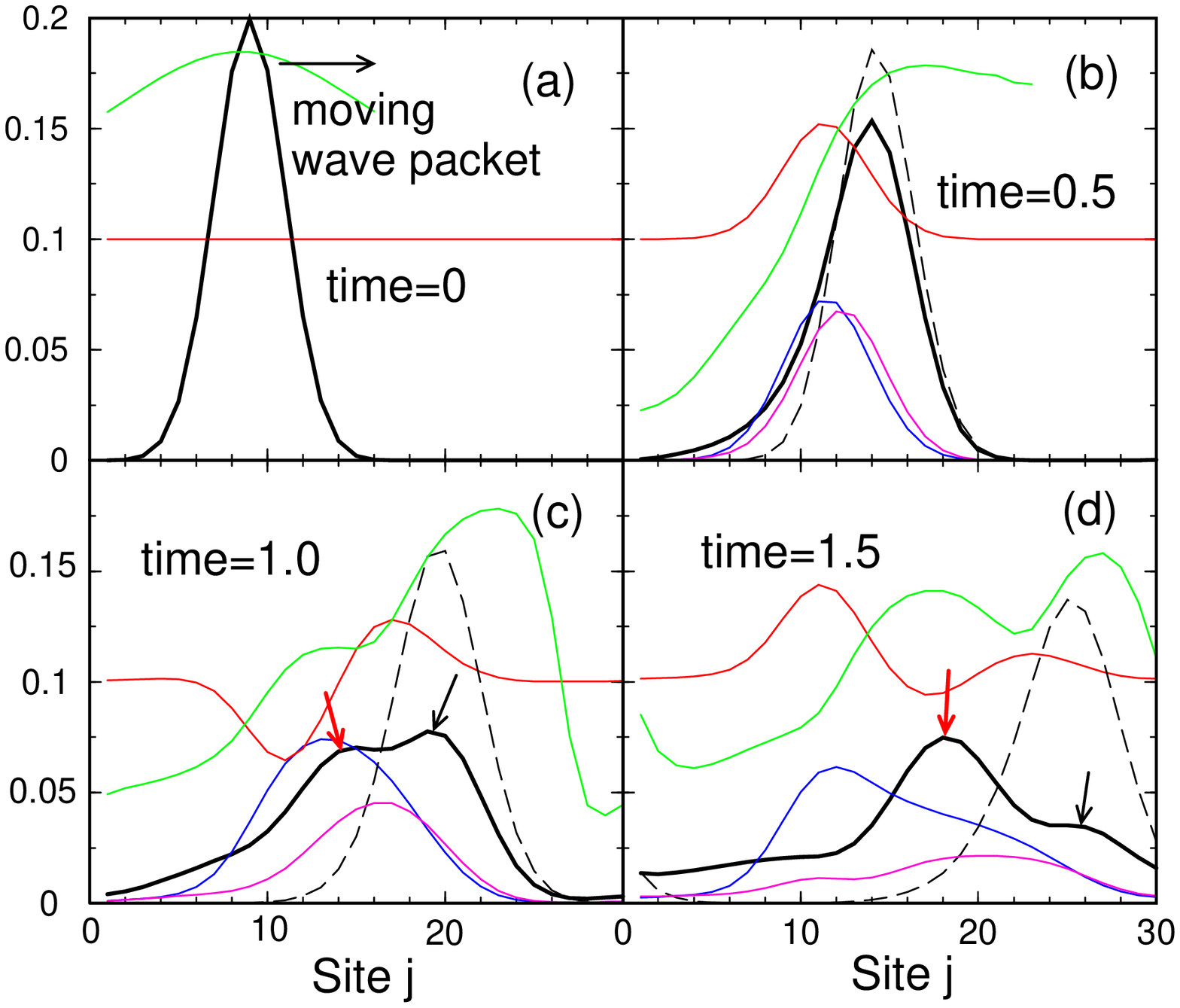,height=12.8cm,width=15cm,angle=0}
} \caption{ Snapshots of the polaron-formation process, for hopping
$t=\omega_0=1$, and $\lambda=0.4$. The calculation is performed on a 30-site
periodic lattice. Time is measured in phonon periods. Black: electron density
$\langle c_j^\dag c_j \rangle$; Blue: phonon density $\langle a_j^\dag a_j
\rangle$; Red: lattice displacement $\langle a_j + a_j^\dag \rangle$; Green:
velocity in units of lattice constant per phonon period; Magenta: el-ph
correlation function $\langle c_j^\dag c_j a_j^\dag a_j \rangle $; Dashed:
free-electron wave packet for reference. For clarity, the origins of the red
and green curves are offset by 0.1 and their values are rescaled by a factor of
0.2 and $0.05/(2\pi)$ respectively. \label{fig:snapshot}} \efig

\bfig  \centerline{\psfig{figure=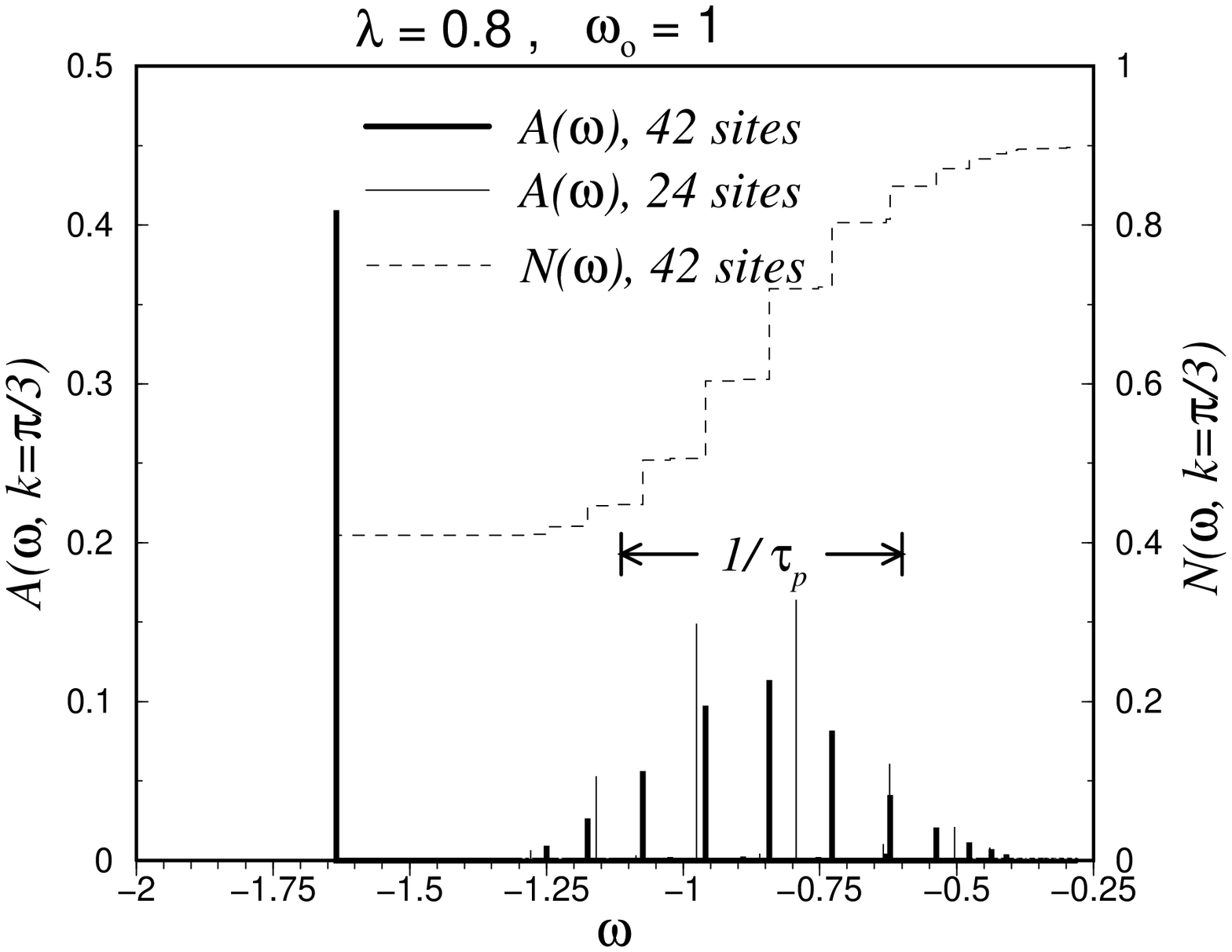,height=13cm,width=16cm,angle=0}}
\caption{Spectral function at weak coupling. $\tau_p$ is the polaron formation
time.}\label{fig:spc1} \efig

\newpage \bfig \centerline{
\psfig{figure=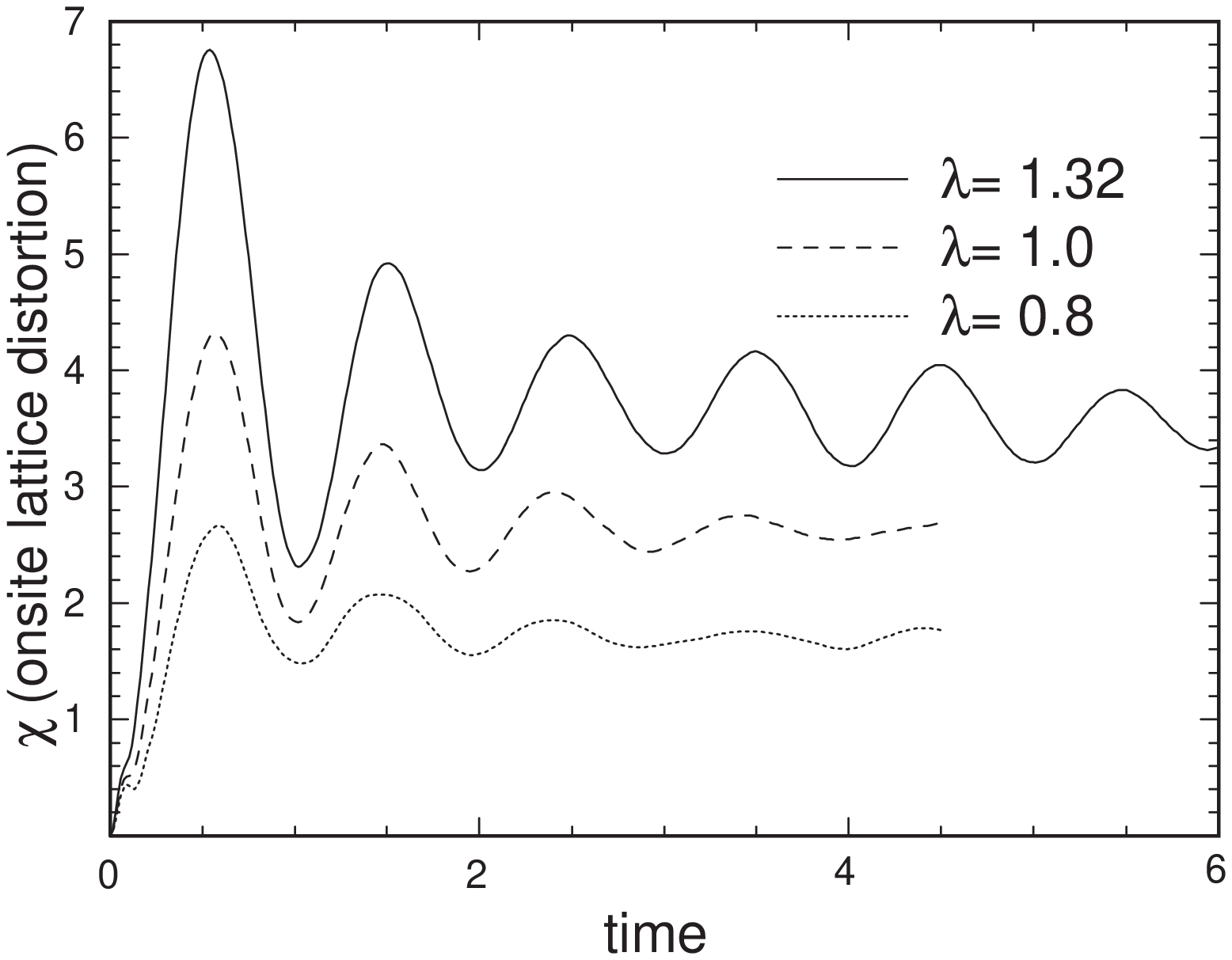,height=13cm,width=16cm,angle=0} } \caption{ The
on-site correlation function $\chi$ as a function of time. The initial energy
of the bare electron (or exciton) is $E_i=-0.7$. When $\lambda=1.0$, the result
most resembles the experimental data \cite{Sugita}. For all curves,
$\omega_0=0.5$ and $t=1$. \label{fig:Sugita}  } \efig

\bfig
 \centerline{\psfig{figure=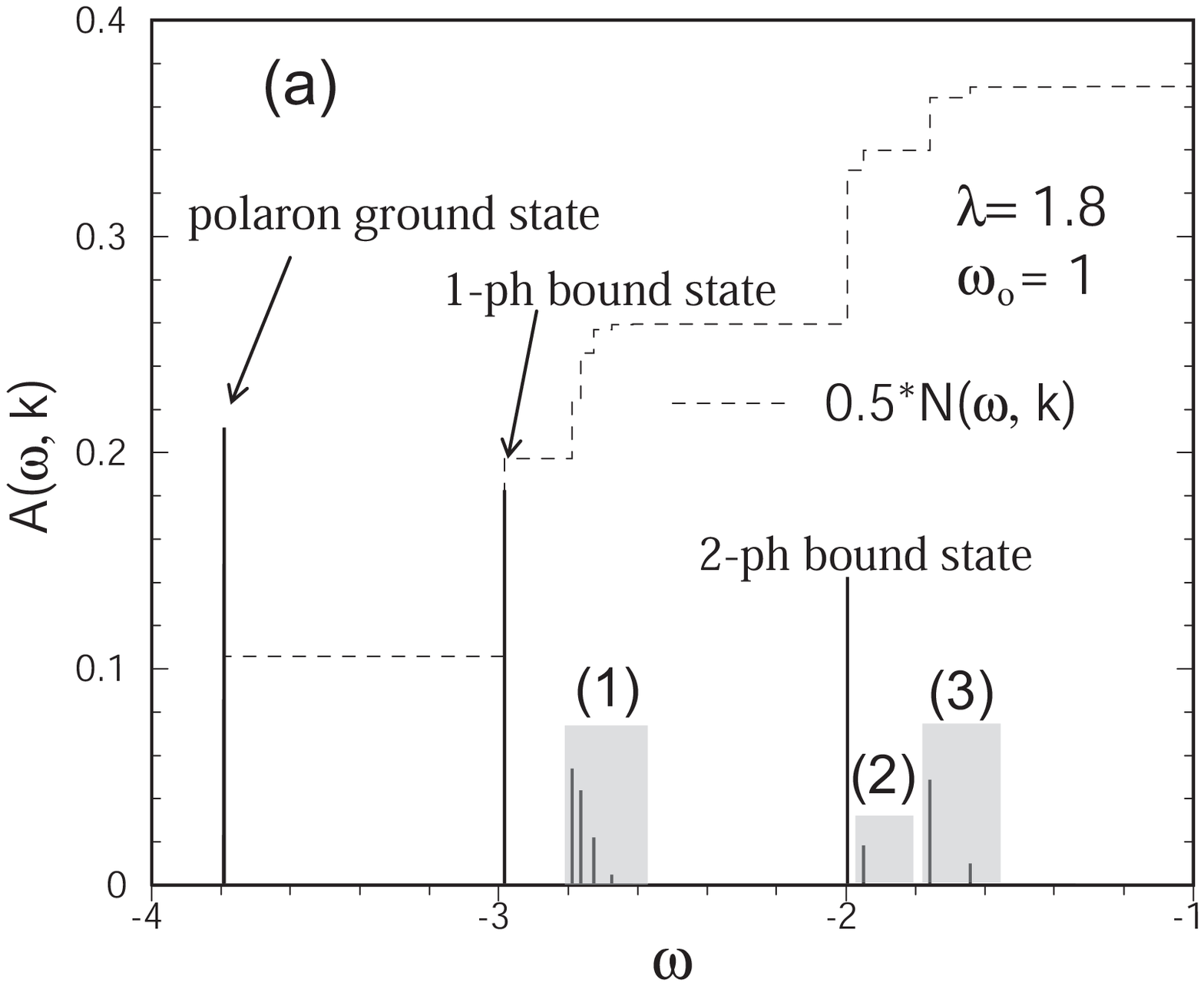,height=11cm,width=14cm,angle=0}}
 \centerline{\psfig{figure=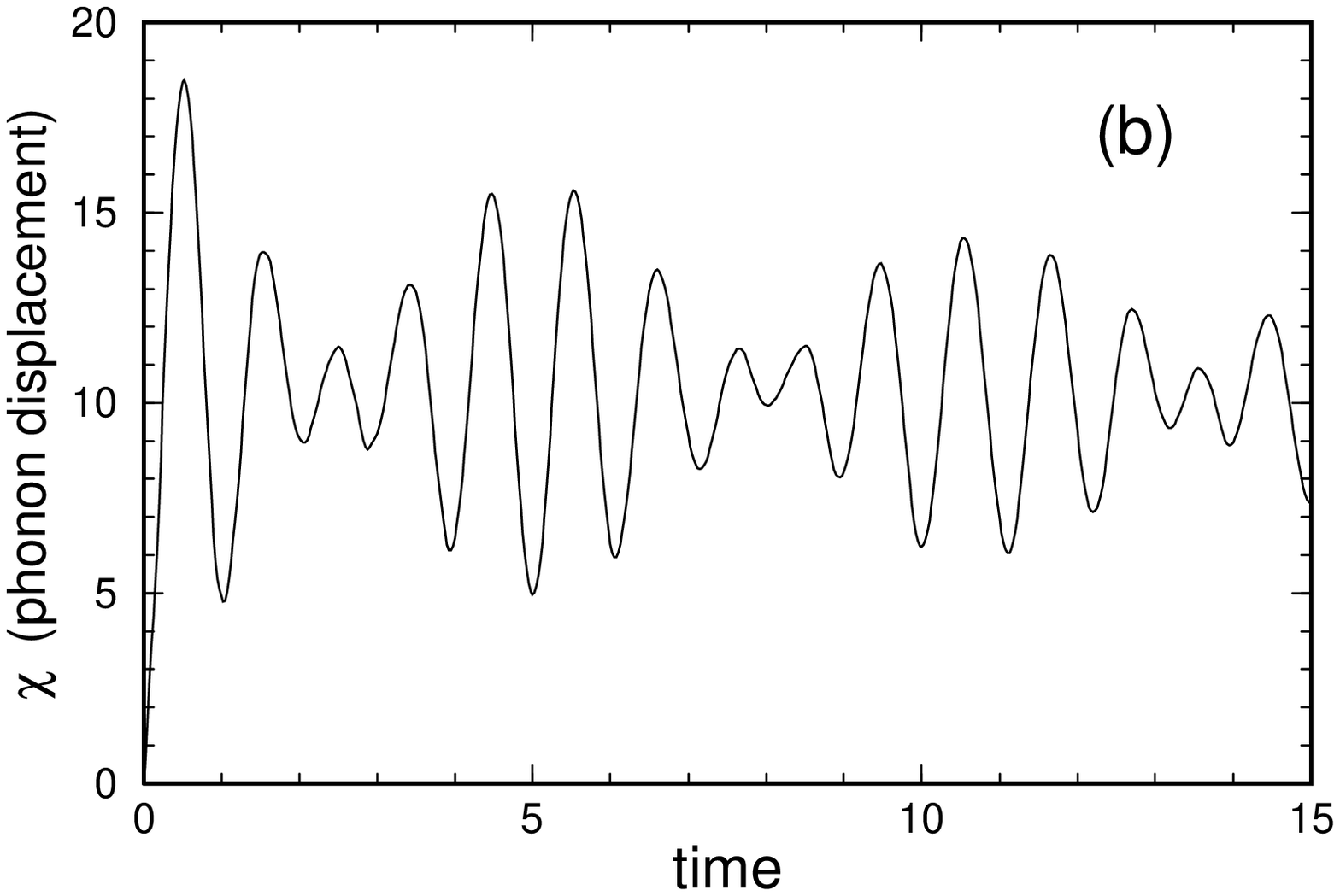,height=8.5cm,width=13.2cm,angle=0} }
 \caption{Panel (a): spectral function at strong coupling. There exist
 three quasiparticle poles. Shaded areas (1) and (2) correspond to continuum
 states (A.2) and (B.2) respectively.  (b): Quantum beat formed by multiple
 quasiparticle poles. } \label{fig:spc2}
 \efig

\bfig \centerline{
 \psfig{figure=Level12Dn.eps,height=7cm,width=16cm,angle=0} } \vskip 2cm \centerline{
 \psfig{figure=Level3Dnx.eps,height=9.5cm,width=12cm,angle=0}}
 \caption{ Eigenvalues of low-lying states as functions of coupling constant in
  1D through 3D. Hopping $t=1$ in all panels.
  In the adiabatic regime in higher dimensions, the ground
  state (thick solid lines) shows a fairly abrupt change in slope.
  In the 3D panel,
  $\psi_1$ and $\psi_4$ are a lightly-dressed quasiparticle state;  $\psi_2$ and
  $\psi_3$ are a heavy polaron state. The dashed lines are the beginning of the
  lowest continua.}
  \label{fig:levels_3D}
 \efig


\begin{references}
\bibitem{otherstates}  This list of states is not complete.
For example, there may be more than one type of single phonon plus
polaron bound state, and the excited states in the
vicinity of the avoided crossings shown in
Fig.~\ref{fig:levels_3D} in 2D and 3D are not included.
\end{references}
\end{document}